\documentclass[12]{iopart}
\usepackage{amsfonts}
\usepackage{epsfig}

\begin{document}

\markboth{Taotao Hu et al.}{Braid groups and entanglement}

\title{Method of constructing braid group representation and entanglement in a $ 9\times9 $  Yang-Baxter sysytem}

\author{Taotao Hu  }
\address{School of Physics, Northeast Normal University,
Changchun 130024, People's Republic of China}
\author{Gangcheng Wang}
\address{School of Physics, Northeast Normal University,
Changchun 130024, People's Republic of China}
\author{Chunfang Sun}
\address{School of Physics, Northeast Normal University,
Changchun 130024, People's Republic of China}
\author{Chengcheng Zhou}
\address{School of Physics, Northeast Normal University,
Changchun 130024, People's Republic of China}
\author{Qingyong Wang}
\address{School of Physics, Northeast Normal University,
Changchun 130024, People's Republic of China}
\author{Kang Xue }
\ead{Xuekang@nenu.edu.cn}
\address{School of Physics, Northeast Normal University,
Changchun 130024, People's Republic of China}

\begin{abstract}
 In this paper we present reducible representation of the $n^{2}$ braid group representation which is constructed
on the tensor product of n-dimensional spaces. Specifically, it is
shown that via a combining method we can construct more $n^{2}$
 dimensional braiding matrix S which satisfy the braid relations.
 By Yang-Baxteraition approach, we derive a $ 9\times9 $ unitary $ \breve{R}$-matrix according to a $ 9\times9 $ braiding S-matrix we have constructed
. The entanglement properties of $ \breve{R}$-matrix is
investigated, and the arbitrary degree of entanglement for
two-qutrit entangled states can be generated via $ \breve{R}$-matrix
acting on the standard basis.
\end{abstract}


 \maketitle
\section{Introduction}
Quantum entanglement is the most surprising nonclassical property of
composite quantum systems that Shr\"{o}dinger singled out many
decades ago as "the characteristic trait of quantum mechanics".
Recently entanglement has become one of the most fascinating topics
in quantum information theory , entanglement is recognized as an
essential resource for quantum processing and quantum
communications\cite{A.K,C.H.L,C.H.B} and it play  a crucial role in
quantum computation\cite{C.H,R.S,S.J}. It is believed that the
protocols based on the entangled states have an exponential speedup
over the classical ones. Besides, in highly correlated states in
condensed-matter systems such as superconductors\cite{S.O,V.O} and
fractional quantum Hall liquids\cite{X.G}, the entanglement serves
as a unique measure of quantum correlations between degrees of
freedom.
 Leveraging the entanglement and using
quantum coherence, certain problems may be solved faster by a
quantum computer than a classical one.

 Recently, it has been revealed that there are natural and profound connections between
quantum computations and braid group theory as well as the
Yang-Baxter equation(YBE)\cite{H.A,L.H,Y.Z,Y.Z.L,J.F,M.N,C.N,R.J}.
During the investigation of the relationships among quantum
entanglement, topological entanglement and quantum computation,
Kauffman and Lomonaco \cite{K.L} have explored the important role of
unitary braiding operators. It is shown that the braid matrix can be
identified as the universal quantum gate \cite{K.L,Z.Y}. This
motivates a novel way to study quantum entanglement based on the
theory of braiding operators, as well as YBE. The first step along
this direction is initiated by Zhang et al \cite{Z.Y}. In
\cite{K.L}, the Bell matrix generating two-qubit entangled states
has been recognized to be a unitary braid transformation. Later on,
an approach to describe Greenberger-Horne-Zeilinger (GHZ) states or
N-qubit entangled states based on the theory of unitary braid
representations has been presented in \cite{Z.Y.G}. Chen and his
co-workers \cite{C.J.L,C.J.X} used unitary braiding operators to
realize entanglement swapping and generate the GHZ states, as well
as the linear cluster states. These literatures introduce the
braiding operators and Yang-Baxter equations to the field of quantum
information and quantum computation. In a very recent work
\cite{C.J.L.L}, it has been found that any pure two-qudit entangled
state can be achieved by a universal Yang-Baxter equation.

  In our paper we present the method of constructing $n^{2}$ dimensional matrix solutions of braid group algebra relation. The paper is organized as follows. In sec II, we present the
reducible representations of $n^{2}$ braid group algebra.
Specifically, more $n^{2}$ dimensional braiding matrix S which
satisfy the braid relations can be obtained by  the combining
method, and we get some well known and some new braiding matrix S.
In sec III, By Yang-Baxteraition approach, we derive a $ 9\times9 $
unitary $ \breve{R}$-matrix according to a $ 9\times9 $ S-matrix we
have constructed. we investigate the entanglement properties of $
\breve{R}$-matrix. It shows that the arbitrary degree of
entanglement for two-qutrit entangled states can be generated via
the unitary matrix $ \breve{R}$ -matrix acting on the standard
basis. The summary is made in the last section.

\section{Method of constructing braiding S-Matrixs}\label{sec2}

In a recent paper\cite{N.M} a reducible representation of the
Temperley-Lieb algebra is constructed on the tensor product of
n-dimensional spaces. In fact Temperley-Lieb algebra is a subalgebra
of braid algebra. Motivated by this, we investigated the methods of
constructing braid representation to get more useful braid
representations conveniently.

We first review the theory of braid groups, Let $B_{n}$ denotes the
braid group on $n$ strands. $B_{n}$ is generated by elementary
braids $\{b_{1},b_{2}, \cdots ,b_{n-1}\}$ with the braid relations,
\begin{eqnarray}\label{braid}
\left\{
\begin{array}
[c]{ll}
b_{i}b_{i+1}b_{i}=b_{i+1}b_{i}b_{i+1} & 1\leq i<n-2\\
& \\
b_{i}b_{j}=b_{j}b_{i} & \left\vert i-j\right\vert \geq2
\end{array}
\right.
\end{eqnarray}
where the notation $b_{i}\equiv b_{i,i+1}$ is used, $b_{i,i+1}$
represents $1_{1}\otimes 1_{2}\otimes 1_{3}\cdots \otimes
S_{i,i+1}\otimes \cdots \otimes 1_{n}$ , and $1_{j}$ is the unit
matrix of the j-th particle.

  By calculation, we get the reducible representations of braiding matrix which is defined by two $n\times n$ matrixs  A and B $\in GL(n,\mathbb{C})$ which all can
also be seen as an $n^{2}$ dimensional vector $\{A^{a}_{b},
B^{c}_{d}\}$ $\in \mathbb{C}^{n}\bigotimes \mathbb{C}^{n}$.

The braiding matrix $S$ can be expressed as

\begin{eqnarray}\label{M}
S^{ab}_{cd}=A^{a}_{d}B^{b}_{c}\in Mat(\mathbb{C}^{n}_{i}\bigotimes
\mathbb{C}^{n}_{i+1})
\end{eqnarray}
where we explicitly write the indices corresponding to the factors
in the tensor product space $\mathcal
{H}$=$\bigotimes^{N}_{1}\mathbb{C}^{n}$. Substituting the relation
into braid relations eq (\ref{braid}), the limited conditions can be
derived. The S in eq (\ref{M}) is a solution of braid relation if
and only if (the detail calculation is given in Appendix A)
\begin{eqnarray}\label{g}
AB=BA,~~~~ namely ~~~~[A,B]=0_{n\times n}
\end{eqnarray}

For example, n=2, in order to get significative result we set that
every row and array of two $2\times 2$ convertible matrix-A and
matrix-B have only one element which is equal to 1 for convenience.
\begin{eqnarray}\label{A}
 A=\left(
  \begin{array}{cc}
    1 & 0 \\
    0 & 1\\
  \end{array}
\right)~~~~~~~~~
 B=\left(
  \begin{array}{cc}
    0 & 1 \\
    1 & 0\\
  \end{array}
\right)
\end{eqnarray}
Substituting eq (\ref{A}) into eq (\ref{M}) we get
\begin{eqnarray}
S=\left(
  \begin{array}{cccc}
    1 & 0 & 0 &0 \\
    0 & 0 & 1 &0\\
    0 & 1 & 0 &0\\
    0 & 0 & 0 &1\\
  \end{array}
\right)~~~~and~~~~
 S=\left(
  \begin{array}{cccc}
    0& 0 & 0 &1 \\
    0 & 1 & 0 &0\\
    0 & 0 & 1 &0\\
    1 & 0 & 0 &0\\
\end{array}
\right)
\end{eqnarray}
the first S we get is the standard swap gate\cite{K.L}, in order to
obtain more useful braiding S-matrix we do the further combination
as follows:
\begin{eqnarray}\label{1}
S=\sum_{i=1}^{2}a_{i}S^{(i)}
\end{eqnarray}

where $S^{(1)}$ and $S^{(2)}$ all have the reducible representations
as eq (\ref{M}), $a_{1}$ and $a_{2}$ are the corresponding
coefficients.
\begin{eqnarray}\label{2}
(S^{(1)})^{ab}_{cd}=(A^{(1)})^{a}_{d}(B^{(1)})^{b}_{c}~~~~(S^{(2)})^{ab}_{cd}=(A^{(2)})^{a}_{d}(B^{(2)})^{b}_{c}
\end{eqnarray}
according to eq (\ref{braid}), eq (\ref{1}) and eq (\ref{2}) we get
when $[A^{(i)},B^{(i)}]=0$, $[A^{(i)},A^{(j)}]=0$,
$[B^{(i)},B^{(j)}]=0$ (i,j=1,2)the constructed S-matrix in eq
(\ref{1}) is a braiding-matrix which satisfy the braid relation eq
(\ref{braid}).

According to the limitation, we set
\begin{eqnarray}\label{A}
 A=\left(
  \begin{array}{cc}
    1 & 0 \\
    0 & 1\\
  \end{array}
\right),~~~~~~
 B=\left(
  \begin{array}{cc}
    0 & i \\
    1 & 0\\
  \end{array}
  \right)~~~~and~~~~~~
C=\left(
  \begin{array}{cc}
    0 & 1 \\
   -i & 0\\
  \end{array}
\right)
\end{eqnarray}
 The coefficient $a_{1}$  and  $a_{2}$ don't have
restriction and  we set them equal to $\frac{ 1}{\sqrt{2}}$ for
convenience. We let $A^{(1)}=B^{(2)}=A$, $B^{(1)}=B$,  $A^{(2)}=C$
and $A^{(1)}=B^{(1)}=A$, $A^{(2)}=B$, $B^{(2)}=C$ respectively,
according to this combining method we can get two $4\times4$ models
as follows:
\begin{eqnarray}
 S=\frac{1}{\sqrt{2}}\left(
  \begin{array}{cccc}
    0& 1& i &0 \\
    1& 0 & 0 &  1\\
    -i & 0 & 0 & i\\
    0 & 1 & -i& 0\\
\end{array}
\right)~~~~and~~~~
 S=\frac{1}{\sqrt{2}}\left(
  \begin{array}{cccc}
    1 & 0 & 0 &i \\
    0 & 1 & 1 &0\\
    0 & 1 & 1 &0\\
    -i & 0 & 0 &1\\
  \end{array}
\right)
\end{eqnarray}
one can see by the combination we obtain two $4\times4$ braiding
model while the first S-matrix is a new braiding model which is
found to be locally equivalent to the DCNOT gate \cite{W.A.G}. This
motivates us to find generalized $n^{2}$ $(n\geq2)$ dimensional
braiding-matrix representation by the combining method. We do the
similar combination as follows:
\begin{eqnarray}\label{3}
S=\sum_{i=1}^{n}a_{i}S^{(i)}
\end{eqnarray}
where $S^{(i)}$ also have the reducible representation
\begin{eqnarray}\label{4}
(S^{(i)})^{ab}_{cd}=(A^{(i)})^{a}_{d}(B^{(i)})^{b}_{c}
\end{eqnarray}
substituting  eq (\ref{3}) and eq (\ref{4}) into eq (\ref{braid} ),
we find $A^{(i)}$ and $B^{(i)}$ subject to the limited conditions as
follows (the detail calculation is given in Appendix A):
\begin{eqnarray}\label{5}
[A^{(i)},B^{(i)}]=0, [A^{(i)},A^{(j)}]=0, [B^{(i)},B^{(j)}]=0 \nonumber\\
(i,j=1,2,3\ldots n)
\end{eqnarray}

coefficients $a_{i}$ are not restricted, when eq (\ref{5}) is
satisfied, S-matrix in eq (\ref{3}) satisfy the braid relation eq
(\ref{braid}). Namely we can obtain more $n^{2}$ dimensional
braiding-matrix representation by this combining method.

\section{A $9\times9$ braiding S-matrix,Yang-Baxterization and Entanglement}\label{sec3}
In section II, we present that we can get arbitrary $n^{2}$
dimensional braiding-matrix representation  by the reducible
representation and the combining method. Now we emphasize on one
$9\times9$ braiding S-matrix we have constructed to investigate it's
application on quantum entanglement.

For n=3, let three  $3\times3$ matrix A, B and C as follows(we
choose $\{|0\rangle,|1\rangle,|2\rangle\}$ as the standard basis),
\begin{eqnarray}
 A=\left(
  \begin{array}{ccc}
    0 & 0 & e^{i\varphi_{1}} \\
    1 & 0 & 0\\
    0 & e^{-i\varphi_{2}}& 0\\
  \end{array}
\right)
 B=\left(
  \begin{array}{ccc}
    0 & 1 & 0 \\
    0 & 0 & e^{i\varphi_{2}}\\
   e^{-ii\varphi_{1}} & 0& 0 \nonumber~~~\\
  \end{array}
\right)
\end{eqnarray}
\begin{eqnarray}
 C=\left(
  \begin{array}{ccc}
    1 & 0 & 0 \\
    0 & 1 & 0\\
    0 & 0& 1\\
  \end{array}
\right)
\end{eqnarray}
here [A,B]=0, [A,C]=0, [B,C]=0 satisfy the eq (\ref{5}), the
parameters $\varphi_{1}$ and $\varphi_{1}$ are both real. we let
$A^{(1)}=B^{(2)}=A$, $A^{(2)}=B^{(1)}=B$ and $A^{(3)}=B^{(3)}=C$,
and we set coefficient $a_{i}$, (i=1,2,3) all equal to 1 for
convenience. By combination, In terms of the standard basis$\{|00\rangle,|01\rangle,|02\rangle,|10\rangle,|11\rangle,|12%
\rangle,|20\rangle,|21\rangle,|22\rangle\}$ we get a $9\times9$
braiding S-matrix as follows:
\begin{eqnarray}\label{6}
S=\left(
  \begin{array}{ccccccccc}
    1 & 0 & 0 &0 & 0 & q_{1} & 0 &q_{1} & 0\\
    0 & 1 & 0 &1 & 0 & 0 & 0 &0 & Q\\
    0& 0 & 1 &0 &q_{2}^{-} & 0 & 1&0 & 0\\
    0 & 1 & 0 &1 & 0 & 0 & 0 &0 & Q\\
    0& 0 & q_{2} &0 & 1 & 0 & q_{2}&0 & 0\\
    q_{1}^{-} & 0 & 0 &0 & 0 & 1 & 0 &1 & 0\\
    0& 0 & 1 &0 & q_{2}^{-}& 0 & 1&0 & 0\\
    q_{1}^{-} & 0 & 0 &0 & 0 & 1 & 0 &1 & 0\\
     0 & Q^{-} & 0 &Q^{-}& 0 & 0 & 0 &0 & 1\\
\end{array}
\right)
\end{eqnarray}
here $q_{1}= e^{i\varphi_{1}}$, $q_{2}=e^{i\varphi_{2}}$,
$Q=q_{1}q_{2}$, and one can easily find that $S^{2}=3S$, $S^{+}=S$.

The usual YBE takes the form \cite{J.L.K}:
\begin{equation}\label{7}
\breve{R}_{i}(x)\breve{R}_{i+1}(xy)\breve{R}_{i}(y)=\breve{R}_{i+1}(y)\breve{R}_{i}(xy)\breve{R}_{i+1}(x).
\end{equation}
The spectral parameters x and y which are related with the
one-dimensional momentum play an important role in some typical
models\cite{C.N}. The asymptotic behavior of $ \breve{R}(x ,
\varphi_{1},\varphi_{2})$ is x-independent, i.e.
$\lim\breve{R}_{i,i+1}(x , \varphi_{1},\varphi_{2})\propto b_{i}$,
where $b_{i}$ are braiding operators, which satisfy the braiding
relations eq (\ref{braid}). From a given solution of the braid
relation $S$, a $\breve{R}(x)$ can be constructed by using the
approach of Yang-Baxterization. Let the unitary Yang-Baxter matrix
take the form,
\begin{equation}\label{8}
\breve{R}(x)=\rho(x) (\mathcal {I} + G(x) S).
\end{equation}

This is a trigonometric solution of YBE, where $\rho(x)$ is a
normalization factor. One can choose appropriate $\rho(x)$ to ensure
that $\breve{R}(x)$ is unitary. Substituting eq (\ref{8}) to eq
(\ref{7}) and according to $S^{2}=3S$, one has
G(x)+G(y)+3G(x)G(y)=G(xy). In addition, the initial condition
$\breve{R}_{i}(x)=I_{i}$ yields G(x=1)=0 and $\rho(x=1)=1$. The
unitary condition (i.e.,
$\breve{R}^{\dag}_{i}(x)=\breve{R}^{-1}(x)=\breve{R}(x^{-})$) can be
tenable only on condition that
$\rho(x)\rho(x^{-})$$(G(x)+G(x^{-})+3G(x)G(x^{-}))$=0. Take account
into these condition, we obtain a set solution of G(x) and
$\rho(x)$,
\begin{equation}
\rho(x)=x,~~~~~~G(x)=-\frac{x-x^{-}}{3x}.\label{9}
\end{equation}
Substituting Eq(\ref{6}), Eq(\ref{9})into Eq (\ref{8}), the unitary
solution of YBE can be obtained as following,
\begin{eqnarray}\label{10}
 \breve{R}_{i}(x , \varphi_{1},\varphi_{2})=\left(
  \begin{array}{ccccccccc}
    b & 0 & 0 &0 & 0 & aq_{1} & 0 &aq_{1} & 0\\
    0 & b & 0 &a & 0 & 0 & 0 &0 & a Q\\
    0& 0 & b &0 &\frac{a}{q_{2}} & 0 & a&0 & 0\\
    0 & a & 0 &b& 0 & 0 & 0 &0 & a Q\\
    0& 0 & aq_{2} &0 & b & 0 & aq_{2}&0 & 0\\
    \frac{a}{q_{1}} & 0 & 0 &0 & 0 & b & 0 &a & 0\\
    0& 0 & a &0 & \frac{a}{q_{2}}& 0 & b&0 & 0\\
    \frac{a}{q_{1}}  & 0 & 0 &0 & 0 & a& 0 &b & 0\\
     0 & \frac{a}{Q} & 0 &\frac{a}{Q} & 0 & 0 & 0 &0 & b\\
\end{array}
\right)
\end{eqnarray}
where a=$x^{-1}-x$, $b=2x+x^{-1}$.
The Gell-Mann matrices, a basis for the Lie algebra SU(3)\cite{wp}, $\lambda_{u}$ satisfy $%
[I_{\lambda},I_{\mu}]=if_{\lambda\mu\nu}I_{\nu}
(\lambda,\mu,\nu=1,\cdot \cdot \cdot ,8)$, where
$I_{\mu}=\frac{1}{2}\lambda_{\mu}$. As a resent paper having done we denote $I_{\lambda}$ by, $I_{\pm}=I_{1}\pm iI_{2}$, $%
V_{\pm}=V_{4}\mp iV_{5}$,$U_{\pm}=I_{6}\pm iI_{7}$,
$Y=\frac{2}{\sqrt{3}}I_{8} $.  we also generate three sets of
realization of $SU(3)$ as:
\begin{eqnarray*}\label{su31}
\left\{
\begin{array}{lll}
I_{\pm}^{(1)}=I_{1}^{\pm}I_{2}^{\mp},~~~U_{\pm}^{(1)}=U_{1}^{\pm}V_{2}^{
\mp},~~~V_{\pm}^{(1)}=V_{1}^{\pm}U_{2}^{\mp},\\
&\\
I_{3}^{(1)}=\frac{1}{3}(I_{1}^{3}-I_{2}^{3})+\frac{1}{2}%
(I_{1}^{3}Y_{2}-Y_{1}I_{2}^{3}),\\
&\\
Y^{(1)}=\frac{1}{3}(Y_{1}+Y_{2})-\frac{2}{3}I_{1}^{3}I_{2}^{3}-\frac{1}{2}
Y_{1}Y_{2};
\end{array}
\right.
\end{eqnarray*}
\begin{eqnarray*}\label{su32}
\left\{
\begin{array}{lll}
I_{\pm}^{(2)}=U_{1}^{\pm}U_{2}^{\mp},~~~U_{\pm}^{(2)}=V_{1}^{\pm}I_{2}^{%
\mp},~~~V_{\pm}^{(2)}=I_{1}^{\pm}V_{2}^{\mp} , \\
& \\
I_{3}^{(2)}=\frac{1}{2}[-\frac{1}{3}(I_{1}^{3}-I_{2}^{3})+\frac{1}{2}%
(Y_{1}-Y_{2})+I_{1}^{3}Y_{2}-Y_{1}I_{2}^{3}], \\
&\\
Y^{(2)}=-[\frac{1}{3}(I_{1}^{3}+I_{2}^{3})+\frac{1}{6}(Y_{1}+Y_{2})+\frac{2}{3%
}I_{1}^{3}I_{2}^{3}+\frac{1}{2}Y_{1}Y_{2}] ;&\\
\end{array}
\right.
\end{eqnarray*}
\begin{eqnarray*}\label{su33}
\left\{
\begin{array}{lll}
I_{\pm}^{(3)}=V_{1}^{\pm}V_{2}^{\mp},~~~U_{\pm}^{(3)}=I_{1}^{\pm}U_{2}^{%
\mp},~~~V_{\pm}^{(3)}=U_{1}^{\pm}I_{2}^{\mp}, \\
&\\
I_{3}^{(3)}=\frac{1}{2}[-\frac{1}{3}(I_{1}^{3}-I_{2}^{3})-\frac{1}{2}%
(Y_{1}-Y_{2})+I_{1}^{3}Y_{2}-Y_{1}I_{2}^{3}],\\
&\\
Y^{(3)}=\frac{1}{3}(I_{1}^{3}+I_{2}^{3})-\frac{1}{6}(Y_{1}+Y_{2})-\frac{2}{3}%
I_{1}^{3}I_{2}^{3}-\frac{1}{2}Y_{1}Y_{2}.\\
\end{array}
\right.
\end{eqnarray*}
We denote $I^{(k)}_{\pm}=I^{(k)}_{1}\pm iI^{(k)}_{2}$, $V^{(k)}_{
\pm}=V^{(k)}_{4}\mp iV^{(k)}_{5}$,$U^{(k)}_{\pm}=I^{(k)}_{6}\pm
iI^{(k)}_{7}$, $Y^{(k)}=\frac{2}{\sqrt{3}}I^{(k)}_{8}$$(k=1,2,3)$.
These realizations satisfy the commutation relation
$[I^{(i)}_{\lambda},I^{(j)}_{\mu}]=i\delta_{ij}f_{
\lambda\mu\nu}I^{(i)}_{\nu}$ $(\lambda,\mu,\nu=1,\cdot \cdot \cdot
,8;i,j=1,2,3)$.

For $i$-th and $(i+1)$-th lattices, $\breve{R}$-matrix can be
expressed in terms of above operators,
\begin{eqnarray*}\label{operator}
\begin{array}{llll}
\breve{R}(x,\varphi_{1},\varphi_{2})=\frac{1}{3}a[I_{+}^{(1)}+I_{-}^{(1)}+Q(V_{-}^{(1)}+U_{+}^{(1)})\\
\\~~~~~~~~~~~~~~~~~~+Q^{-1}(U_{-}^{(1)}+ V_{+}^{(1)})+I_{+}^{(2)}+I_{-}^{(2)} \\
\\~~~~~~~~~~~~~~~~~~+q_{1}(V_{+}^{(2)}+U_{-}^{(2)})+q_{1}^{-1}(V_{-}^{(2)}+U_{+}^{(2)}) \\
\\~~~~~~~~~~~~~~~~~~+I_{+}^{(3)}+I_{-}^{(3)} +q_{2}(V_{+}^{(3)}+U_{-}^{(3)})\\
\\~~~~~~~~~~~~~~~~~~+q_{2}^{-1}(V_{-}^{(3)}+U_{+}^{(3)})]+\frac{b}{3}(I\otimes I).
\end{array}
\end{eqnarray*}
So the whole tensor space $\mathbb{C}^{3}\otimes\mathbb{C}^{3}$ is
completely decomposed.i.e.
$\mathbb{C}^{3}\otimes\mathbb{C}^{3}=\mathbb{C}^{3}\oplus
\mathbb{C}^{3}\oplus \mathbb{C}^{3}$. In addition, each block of
$\breve{R}$-matrix can be represented by fundamental representation
of SU(3) algebra.

According to the condition
$\breve{R}^{\dag}_{i}(x)=\breve{R}^{-1}(x)$  one can get
$x^{\ast}=-x$, so we can introduce a new parameter with
x=$e^{i\theta}$, and $\theta$ may be related with entanglement
degree. When one acts $\breve{R}(\theta,\varphi_{1},\varphi_{2})$ on
the separable state $|mn\rangle$ , he yields the following family of
states
$|\psi\rangle_{mn}=\sum_{ij=00}^{22}\breve{R}^{ij}_{mn}|mn\rangle$(m,n=0,1,2).
For example, if m=0 and n=0,
\begin{equation}\label{the}
|\psi\rangle_{00}=\frac{1}{3}(b|00\rangle+aq_{1}^{-1}|12\rangle+
aq_{1}^{-1}|21\rangle)
\end{equation}
In Ref. \cite{S.A}, the generalized concurrence (or the degree of
entanglement \cite{S.H.I}) for two qudits is given by
\begin{equation}\label{che}
\mathcal {C}=\sqrt{\frac{d}{d-1}(1-I_{1})}
\end{equation}
where
$I_{1}$=$Tr[\rho_{A}^{2}]$=$Tr[\rho_{B}^{2}]$=$|\kappa_{0}|^{4}$+$|\kappa_{1}|^{4}$+$\cdots+|\kappa_{d-1}|^{4}$,
$\rho_{A} $ and  $\rho_{B}$ are the reduced density matrices for the
sub-systems, and $\kappa_{j}$'s ($j=0,1,\ldots,d-1$) are the Schmidt
coefficients. Then we can obtain the generalized concurrence of the
state $|\psi\rangle_{00}$ as
\begin{eqnarray}\label{con}
\mathcal
{C}&=&\sqrt{\frac{3}{2}(1-\frac{1}{81}|2x+x^{-}|^{4}-\frac{2}{81}|x-x^{-}|^{4})}\nonumber\\
&=&\frac{2\sqrt{2}}{3}|sin\theta|\sqrt{2cos^{2}\theta +1}
\end{eqnarray}
one can find that when $\theta= \frac{\pi}{3}$, the state
$|\psi\rangle_{00}$ becomes the maximally entangled state of tow
qutrits as state $|\psi\rangle_{00}=\frac{1}{\sqrt{3}}%
(e^{i\frac{\pi}{6}}|00\rangle-iq_{1}^{-1}|12\rangle-iq_{1}^{-1}|21\rangle)$.
In general, if one acts the unitary Yang-Baxter matrix
$\breve{R}(x)$ on the
basis $\{|00\rangle,|01\rangle,|02\rangle,|10\rangle,|11\rangle,|12%
\rangle,|20\rangle,|21\rangle,|22\rangle\}$, he will obtain the same
 generalized concurrence
 as Eq(\ref{con}). It is easy to check that the generalized concurrence
ranges from 0 to 1 when the parameter $\theta$ runs from 0 to $\pi$.
But for $\theta \in [0,\pi]$, the generalized concurrence is not a
monotonic function of $\theta$. And when $x=e^{i\frac{\pi}{3}}$, he
will generate nine complete and orthogonal maximally entangled
states for two qutrits. The QE doesn't dependent on the parameters
$\varphi_{1}$ and $\varphi_{2}$. So one can verify that parameter
$\varphi_{1}$ and $\varphi_{2}$ may be absorbed into a local
operation.
\section{Summary}\label{sec4}
In this paper, we have presented the reducible representation of
braid group algebra, Specifically that by the further combining
method we can get more  $n^{2}$ dimensional braiding S-matrixs and
we obtain some well known and new braiding models. According to a $9
\times 9$ braiding $S$-matrix which we have constructed satisfing
the braiding
 relations we derived a unitary
$\breve{R})$-matrix via Yang-Baxterization.  We show that the
arbitrary degree of entanglement for two-qutrit entangled states can
be generated via the unitary $\breve{R}$ matrix acting on the
standard basis.

\section*{Acknowledgments}

This work was supported by NSF of China (Grant No. 10875026).

\appendix

\section{}
The two limited conditions in Sec.2 will be calculated in detail as
follows,

If we substitute $S^{ab}_{cd}=A^{a}_{d}B^{b}_{c}$ into the braid
relation in Eq(\ref{braid})(i.e.
$S_{12}S_{23}S_{12}=S_{23}S_{12}S_{23}$)
\begin{eqnarray}
[S_{12}S_{23}S_{12}]^{abc}_{edf}&=&[S_{12}]^{abc}_{ijk}[S_{23}]^{ijk}_{\alpha\beta\gamma}[S_{12}]^{\alpha\beta\gamma}_{def}\nonumber\\
&=& S^{ab}_{ij}S^{jc}_{\beta f}S^{i\beta}_{de}\nonumber\\
&=& A^{a}_{j}B^{b}_{i}A^{j}_{f}B^{c}_{\beta}A^{i}_{e}B^{\beta}_{d}\nonumber\\
&=& (BA)^{b}_{e}(A^{2})^{a}_{f}(B^{2})^{c}_{d} \label{appeqn1}
\end{eqnarray}
\begin{eqnarray}
[S_{23}S_{12}S_{23}]^{abc}_{edf}&=&[S_{23}]^{abc}_{ijk}[S_{12}]^{ijk}_{\alpha\beta\gamma}[S_{23}]^{\alpha\beta\gamma}_{def}\nonumber\\
&=& S^{bc}_{jk}S^{aj}_{d \beta }S^{\beta k}_{ef}\nonumber\\
&=& A^{b}_{k}B^{c}_{j}A^{a}_{\beta}B^{j}_{d}A^{\beta}_{f}B^{k}_{e}\nonumber\\
&=& (AB)^{b}_{e}(A^{2})^{a}_{f}(B^{2})^{c}_{d}\label{appeqn2}
\end{eqnarray}
according to Eq(\ref{appeqn1}) and Eq(\ref{appeqn2}), one can see if
AB=BA, namely $[A,B]=0$, the braid relation
$S_{12}S_{23}S_{12}=S_{23}S_{12}S_{23}$ holds.

substitute $S=\sum_{i=1}^{n}a_{i}S^{(i)}$ into
$[S_{12}S_{23}S_{12}]^{abc}_{edf}$ and
$[S_{23}S_{12}S_{23}]^{abc}_{edf}$ respectively ,one has

\begin{eqnarray}
[S_{12}S_{23}S_{12}]^{abc}_{edf}&=&\sum_{ghl}^{n}a_{g}a_{h}a_{l}[S_{12}^{(g)}S_{23}^{(h)}S_{12}^{(l)}]^{abc}_{edf}\nonumber\\
&=& \sum_{ghl}^{n}a_{g}a_{h}a_{l}(S^{(g)})^{ab}_{ij}(S^{(h)})^{jc}_{\beta f}(S^{(l)})^{i\beta}_{de}\nonumber\\
 &=&
\sum_{ghl}^{n}a_{g}a_{h}a_{l}
(B^{(g)}A^{(l)})^{b}_{e}(A^{(g)}A^{(h)})^{a}_{f}(B^{(h)}B^{(l)})^{c}_{d}\label{appeqn3}
\end{eqnarray}
\begin{eqnarray}
[S_{23}S_{12}S_{23}]^{abc}_{edf}&=&\sum_{\lambda\mu\nu}^{n}a_{\lambda}a_{\mu}a_{\nu}[S_{23}^{(\lambda)}S_{12}^{(\mu)}S_{23}^{(\nu)}]^{abc}_{edf}\nonumber\\
&=& \sum_{\lambda\mu\nu}^{n}a_{\lambda}a_{\mu}a_{\nu}(S^{(\lambda)})^{bc}_{jk}(S^{(\mu)})^{aj}_{d \beta }(S^{(\nu)})^{\beta k}_{ef}\nonumber\\
 &=&
\sum_{\lambda\mu\nu}^{n}a_{\lambda}a_{\mu}a_{\nu}
(A^{(\lambda)}B^{(\nu)})^{b}_{e}(A^{(\mu)}A^{(\nu)})^{a}_{f}(B^{(\lambda)}B^{(\mu)})^{c}_{d}\label{appeqn4}
\end{eqnarray}
here $(g,h,l=1,2,3\cdots n)$ and $(\lambda,\mu,\nu=1,2,3\cdots n)$
respectively. So we can let $g=\nu$, $h=\mu$ and $l=\lambda$, then
according to Eq(\ref{appeqn3}) and Eq(\ref{appeqn4}), we limit
$A^{\lambda}B^{\nu}=B^{\nu}A^{\lambda}$,
$A^{\nu}A^{\mu}=A^{\mu}A^{\nu}$, and
$B^{\mu}B^{\lambda}=B^{\lambda}A^{\mu}$,
$(\lambda,\mu,\nu=1,2,3\cdots n)$. Under this limited condition, the
Eq(\ref{appeqn3}) is equal to Eq(\ref{appeqn4}), namely when the
limited condition Eq(\ref{5}) is satisfied the braid relation
$S_{12}S_{23}S_{12}=S_{23}S_{12}S_{23}$ holds.

\section*{}


\begin{thebibliography}{99}
\bibitem{A.K}A. K. Ekert, {\it Phys. Rev. Lett} {\bf 67}  (1991) 661.
\bibitem{C.H.L} C. H.
Bennettand S. J. Wiesner, {\it Phys. Rev. Lett} {\bf 69} (1992)
2881.
\bibitem{C.H.B}C. H. Bennett, G. Brassard, C. Cr¨¦peau, R.
Jozsa, A. Peres, and W. K. Wootters, {\it Phys. Rev. Lett} {\bf 70}
 (1993) 1895.
\bibitem{C.H} C. H. Bennett and D. P. Divincenzo, {\it Nature } {\bf 404}  (2000) 247.
\bibitem{R.S} R. Raussendorf and H. J. Briegel, {\it Phys. Rev. Lett} {\bf 86} (2001) 5188.
\bibitem{S.J} S.-S. Li, G.-L. Long, F.-S. Bai, S.-L. Feng, and H.-Z. Zheng,
Proc. Natl. Acad. Sci. U.S.A. {\bf 98} (2001) 11847.
\bibitem{S.O} S. Oh and J.
Kim, {\it Phys. Rev. B} {\bf 71}  (2005) 144523.
\bibitem{V.O} V. Vedral, {\it New J. Phys} {\bf 6}
 (2004) 102.
\bibitem{X.G} X. G. Wen, {\it Phys. Lett. A } {\bf 300} (2002) 175.
\bibitem{H.A} H. A. Dye, Unitary solutions to the Yang-Baxter Equation in Dimension
Four,Quant.Inf.Proc. {\bf 2} (2003£©117-150.Arxiv:quant-ph/0211050.
\bibitem{L.H} L.H. Kauffman and S.J Lomonaco Jr.,Braiding Operators are universal Quantum Gates,{\it New J. Phys} {\bf 6 }(2004)
134. Arxiv:quant-ph/0401090
\bibitem{Y.Z} Y.Zhang, L.H. Kauffman and M.L. Ge, Universal quantum
Gate, Yang-Baxterization and Hamiltonian. Int.J.Quant.Inform.3 no.4
(2005) 669-678.Arxiv:quant-ph/0412095
\bibitem{Y.Z.L} Y.Zhang, L.H. Kauffman and M.L. Ge,
Yang-Baxterization, Universal quantum Gate and Hamiltonians.
Quant.Inf.Proc. {bf 4}£¨2005£©159-197. Arxiv:quant-ph/0502015.
\bibitem{J.F} J.Franko, E.C. Rowell and Z. wang, Extraspecial
2-Groups and Images of Braid Group Representtations. J. knot Theory
Ramifications, {\bf 15}(2006) 413-428. Arxiv: math.RT/0503435.
\bibitem{M.N} M. Nielsen and I.Chuang, Quantum Computation and
Quantum Information(Cambridge University Press) (1999).
\bibitem{C.N} C.N. Phys. Rev. Lett. 19, 1312(1967); C.N. {it\ Phys.
Rev} {\bf 168} (1968) 1920.
\bibitem{R.J}R.J.Baxter, Partition Function of the Eight-vertex
Lattice Model, Annals Phys. {\bf 70 }(1972) 193-228.
\bibitem{K.L} Kauffman L H and
Lomonaco (Jr) S J  {it\ New J. Phys} {\bf 6 }£¨2004£©134.
\bibitem{Z.Y} Zhang Y, Kauffman L H
and Ge M L  Int. J. Quantum Inf. {\bf 3 } (2005) 669.
\bibitem{Z.Y.G}Zhang Y and Ge M L
Quant. Inf. Proc. {\bf 6 }(2007) 363 Zhang Y, Rowell E C, Wu Y S,
Wang Z H and Ge M L  [quant-ph] (2007)
\bibitem{C.J.L} Chen J L, Xue K and Ge M L
{\it Phys. Rev. A} {\bf 76}  (2007) 042324.
\bibitem{C.J.X} Chen J L, Xue K and Ge M L  Ann.
Phys. {\bf 323} (2008) 2614.
\bibitem{C.J.L.L} Chen J L, Xue K and Ge M L  [quant-ph] (2008).
\bibitem{N.M} P. P. Kulish, N. Manojlovic and Z. Nagy  J.
Math. Phys. {\bf 49} (2008) 023510.
\bibitem{W.A.G} Gangcheng Wang, Kang Xue, Chunfeng Wu, He Liang and C
H oh  {it\ J. Phys. A:Math Theor.} {\bf 42 } (2009)  125207.
\bibitem{J.L.K} Chen J L, Xue K, and Ge M L {\it Phys. Rev. A} {\bf 76} (2007) 042324.
\bibitem{wp}W. Pfeifer, The Lie Algebras su(N) An Introduction,
Birkhauser Verlag(2003).
\bibitem{S.A} S. Albeverio and S. M. Fei. J. Opt. B: Quantum
Semiclass. Opt. {\bf 3} (2001) 223.
\bibitem{S.H.I}S.Hii and W. K. Wootters. {\it Phys. Rev. Lett} {\bf 78}
(1997) 5022; W.K. Wooters. {\it Phys. Rev. Lett} {\bf 80}  (1998)
2245.
\bibitem{berry} M. V. Berry, Proc. R. Soc. London, Ser. {\bf{A 392}} (1984) 45.



\end{thebibliography}
\end{document}